\documentstyle[12pt,epsf]{article}
\makeatletter
\@addtoreset{equation}{section}

\makeatother

\begin{document}

\begin{flushright}
Dec 2001

KEK-TH-795
\end{flushright}

\begin{center}

\vspace{5cm}

{\Large Charged Tachyons and Gauge Symmetry Breaking}

\vspace{2cm}

Takao Suyama \footnote{e-mail address : tsuyama@post.kek.jp}

\vspace{1cm}

{\it Theory Group, KEK}

{\it Tsukuba, Ibaraki 305-0801, Japan}

\vspace{4cm}

{\bf Abstract} 

\end{center}

We discuss the condensation of charged tachyons in the heterotic theory on the Kaluza-Klein 
Melvin background. 
The arguments are based on duality relations which are expected to hold from the adiabatic 
argument. 
It is argued that in many cases the rank of the gauge group is not changed by the 
condensation, as opposed 
to naive expectations. 

\newpage

\vspace{1cm}

\section{Introduction}

\vspace{5mm}

To discuss properties of a theory, one has to find a stable vacuum of the theory. 
Thus it is important to investigate the (in)stability of vacua of the 
theory. 
In string theory, there are infinitely many perturbative vacua, and some of them are 
obviously unstable since the theories on such vacua contain tachyons in their mass spectra. 
In general, tachyons are not always excluded by the consistency conditions of the 
worldsheet theory, 
and in fact, there are many non-supersymmetric theories with tachyons. 
Thus it is necessary to consider how to understand such vacua. 

String theories which contain tachyons may look strange. 
However, recent research reveals that tachyons in the open string sector can be regarded as 
something like Higgs bosons \cite{open} (See also e.g. \cite{open2} and references therein). 
There is a nontrivial potential for the tachyons, and the tachyonic instability in the 
original 
vacuum is just due to the fact that the theory is defined around a maximum of the potential. 
The potential has global minima which correspond to stable vacua. 
These vacua are understood as states in which the tachyons condense and have nonzero vevs. 
The open string tachyons appear on unstable D-brane systems, and the tachyon condensation 
describes the decay of such D-brane systems. 
An important lesson we have learned from the open string tachyon condensation is that the 
existence of tachyons would be just a signal of an instability, not the inconsistency of the 
theory. 

It would be expected that tachyonic vacua in closed string theories can also be understood 
in the same spirit. 
An important step toward this direction was made by making a conjecture which relates the 
tachyonic 
instability of Type 0A theory and the instability of the Kaluza-Klein Melvin background 
\cite{CostaGutperle} (below we will call it the KK-Melvin, for short). 
The KK-Melvin is a ``twisted'' compactification of the flat spacetime on $S^1$. 
String theories on the KK-Melvin have been studied in \cite{RussoTseytlin}. 
It was shown that the KK-Melvin is stabilized by an 
instanton effect \cite{instanton}\cite{instanton2} which leads the KK-Melvin to the 
supersymmetric background, i.e. ${\bf R}^{1,n}\times S^1$. 
This leads one to the expectation that Type 0A theory would have a stable ground state which 
corresponds to Type IIA theory with the maximal supersymmetry restored. 
Similar argument was applied to the non-supersymmetric heterotic theories 
\cite{Suyama1}\cite{Suyama2}. 
There are also related works on the KK-Melvin, its generalizations and string theories on 
them \cite{flux-1}-\cite{twistS1}. 

There are other non-supersymmetric backgrounds of string theory, i.e. non-supersymmetric 
orbifolds which are studied in the context of the tachyon condensation. 
They were studied in \cite{APS}, and more deeply investigated recently in 
\cite{Vafa}\cite{c-Th} 
from the point of view of the worldsheet theory. 
Moreover, there is a proposal for the tachyon potential for closed string tachyons \cite{DV}. 

In this paper, we would like to investigate the heterotic theory on the KK-Melvin, by using 
duality relations to other theories, which was briefly discussed in \cite{Suyama2}. 
Our main interest is the gauge symmetry breaking induced by the condensation of charged 
tachyons: 
In many cases, there are tachyons which form a nontrivial representation of the gauge group. 
As argued in \cite{Suyama3}, the condensation of such tachyons would break the gauge 
symmetry and reduce the rank of the gauge group. 
On the other hand, the rank is severely restricted for the theory to be consistent. 
For example, in the supersymmetric heterotic theories in ten dimensions, the rank must equal 
to sixteen to cancel the gauge and the gravitational anomaly. 
Therefore one might worry that the tachyon condensation would spoil the consistency of the 
theory. 
We will argue that at least in some situations it is not the case, and show how such disaster 
can be avioded. 

The organization of this paper is as follows. 
In section \ref{dual}, we discuss duality relations of string theories on the KK-Melvin, 
based on the adiabatic argument \cite{adiabatic}. 
The appearance of tachyonic modes in the heterotic theories on the KK-Melvin is explicitly 
shown in section \ref{spectrum}. 
We consider a dual picture in M-theory in section \ref{M}, and find 
an interesting relation between the tachyonic instability and the dielectric effect 
\cite{Myers}. 
We discuss, in section \ref{string}, dual string theories for several situations to deduce 
the fate 
of the gauge symmetry after the tachyon condensation. 
Section \ref{discuss} is devoted to discussions.

\vspace{1cm}

\section{String duality on Melvin background} \label{dual}

\vspace{5mm}

The duality relations among all the five string theories and the eleven-dimensional 
supergravity 
enable us to investigate many nonperturbative aspects of the theories \cite{Pol}. 
The strong coupling behavior of a theory can be discussed by considering the weak coupling 
behavior of its dual theory. 
There are many pieces of nontrivial evidence for the existence of the dual pairs. 
However, the arguments heavily depend on the existence of the extended supersymmetry, and 
it becomes hard to show such duality relations when there are less number of supersymmetries 
and, of course, when there is no supersymmetry at all. 

There is a nice argument \cite{adiabatic} which enables one to produce a new dual pair from 
a known one. 
The prescription is as follows: 
Suppose that a theory $A$ and a theory $B$ are dual to each other. 
Consider a symmetry $h_A$ of $A$. 
The duality relation ensures that the corresponding symmetry $h_B$ exists in $B$. 
In general, the orbifold $A/h_A$ is not dual to the orbifold $B/h_B$. 
In fact, a counterexample for such an expectation is shown in \cite{adiabatic}. 
A dual pair can be obtained by first compactifying both theories on, for example, $S^1$, and 
then orbifolding them by $h_{A(B)}\sigma_{1/2}$, where $\sigma_{1/2}$ is the shift operator 
along the $S^1$ by a half circumference. 
That is, the theory $A$ on $S^1$ twisted by $h_A\sigma_{1/2}$ is dual to the 
theory $B$ on $S^1$ twisted by $h_B\sigma_{1/2}$. 
To see whether the duality relation holds, let the radius of the $S^1$ be very large. 
Then the effect of the orbifolding will be negligible 
and the local physics of each orbifold is governed by the original theory $A$ or $B$. 
Thus the orbifolds are dual to each other at least locally. 
Moreover, they are dual to each other globally, since they can be understood as an example 
of a pair related by the fiberwise duality. 
An important point of the above construction of dual pairs is that the discrete group 
($h_{A(B)}\sigma_{1/2}$ in the above case) acts freely. 
In fact, $h_A$ alone does not always act freely on $A$, and the simple orbifold $B/h_B$ 
cannot 
always be the dual of $A/h_A$. 

This prescription has been employed to obtain dual pairs with less number of supersymmetries 
\cite{lessSUSY}. 
See also \cite{unify}. 
Moreover, it has also been applied to the cases in which there is no supersymmetry 
\cite{Harvey}, 
to discuss properties of non-supersymmetric theories similar to the one in \cite{KKS} to all 
orders in 
perturbation expansion as well as nonperturbatively. 

\vspace{2mm}

In our previous paper \cite{Suyama2}, we have assumed a duality relation between Type I 
theory and the heterotic theory both on the KK-Melvin, to argue the strong 
coupling behavior of the latter theory, in spite of the absence of 
supersymmetry. 
This is based on a similar argument to the adiabatic argument explained above. 
Since the KK-Melvin is locally flat, the local physics of the former theory should be dual 
to that of the latter one. 
In addition, the duality transformation of the spacetime fields in the supergravity tells us 
that the KK-Melvin 
in the heterotic theory corresponds to the same KK-Melvin in Type I 
theory. 
Therefore the above two theories would be dual to each other. 

Indeed, this is the case to which the adiabatic argument can be applied. 
It has been shown \cite{Suyama1}\cite{Suyama2} that the heterotic theory on 
the KK-Melvin with special values of parameters (and some patterns of Wilson lines) can be 
reinterpreted as an orbifold. 

To make the statement more definite, let us write down the metric of the KK-Melvin, 
\begin{equation}
ds^2 = \eta_{\mu\nu}dx^\mu dx^\nu + dr^2 + r^2(d\theta+qdy)^2 + dy^2, 
\end{equation}
where $\mu,\nu=0,\cdots,6$. 
We have used the polar coordinates $r,\theta$ for the 7-8 plane. 
The $y$-direction is compactified on $S^1$ with the radius $R$. 
There is a real parameter $q$ which represents the nontriviality of the global structure of 
the spacetime. 
The parameter $q$ has a periodicity with the period $2/R$ when there exist spacetime 
fermions. 

The one-loop partition function of the heterotic theory on the KK-Melvin can be obtained 
explicitly, 
\begin{eqnarray}
Z_q(A^I_Y) &=& \int \frac{d^2\tau}{\tau_2}\tau_2^{-4}|\eta(\tau)|^{-12}
         \sum_{m,n\in {\bf Z}}\exp\left[-\frac{\pi R^2}{\alpha'\tau_2}|n+m\tau|^2\right]
         \bigl|Z^{1+2mqR}_{1+2nqR}(\tau)\bigr|^{-2}
               \nonumber  \\
& & \exp\left[\pi imn\left\{(qR)^2-\sum_{I=1}^{16}(A^I_YR)^2\right\}\right]
    e^{2\pi imqR}Z^{(m,n)}_L(\tau){Z^{1+mqR}_{1+nqR}(\tau)^*}^4, 
\end{eqnarray}
where $A^I_Y$ is a Wilson line put on the $y$-direction. 
See \cite{Suyama2} for the details of the expressions. 
This partition function can be rewritten as that of an orbifold, if one chooses $q=1/R$ and 
an appropriate Wilson line. 
The orbifold is the heterotic theory on $S^1$ twisted by the operator 
$(-1)^{F_s}\gamma_\delta\sigma_{1/2}$, where $F_s$ is the spacetime fermion number and 
$\gamma_{\delta}$ is a rigid gauge transformation which is determined by the choice of the 
Wilson line. 
This is a right orbifold theory to apply the adiabatic argument, and the dual theory should 
be the Type I theory on $S^1$ twisted by the operator 
$(-1)^{F_s}\gamma'_\delta\sigma_{1/2}$, where $\gamma'_\delta$ is the corresponding gauge 
transformation in Type I theory. 
One can easily see that the closed string sector of the dual orbifold is just that of Type I 
theory on the KK-Melvin with the same value of $q$. 
Therefore, since the consistency of the theory determines its open string sector, it is 
concluded that the dual of the heterotic theory on the KK-Melvin is Type I theory on the 
same KK-Melvin, as expected. 

\vspace{2mm}

The above relation between the KK-Melvin and the freely-acting orbifold can be found without 
one-loop calculations. 
Consider the worldsheet action of the heterotic theory on the KK-Melvin with a Wilson line, 
\begin{eqnarray}
S &=& \frac1{4\pi\alpha'}\int d^2\sigma\left\{\eta_{\mu\nu}\partial_aX^\mu\partial_aX^\nu
     +|\partial_aX+iq\partial_aYX|^2+(\partial_aY)^2\right\} 
        \nonumber \\
  & & \hspace*{-1cm}
      +\frac i{\pi}\int d^2\sigma\ S^{r\dag}\left(\partial_++\frac i2q\partial_+Y\right)S^r
      +\frac i{\pi}\int d^2\sigma\ 
        \sum_{I=1}^{16}\lambda^{I\dag}\left(\partial_--iA^I_Y\partial_-Y\right)\lambda^I,
\end{eqnarray}
where $r=1,\cdots,4$. 
The right-moving fermions $S^r$ are the Green-Schwarz fermions while the left-moving 
fermions $\lambda^I$ are the RNS fermions. 
This action can be reduced to the free action by the field redefinitions 
\begin{eqnarray}
&& X = e^{-iqY}\tilde{X} \nonumber, \\
&& S^r = e^{-\frac i2qY}\tilde{S}^r, 
         \label{tilde}          \\
&& \lambda = e^{iA^I_YY}\tilde{\lambda}^I. \nonumber 
\end{eqnarray}
The new fields satisfy the twisted boundary conditions 
\begin{eqnarray}
&& Y(\sigma+2\pi) = Y(\sigma) + 2\pi wR, \nonumber \\
&& \tilde{X}(\sigma+2\pi) = e^{2\pi iwqR}\tilde{X}(\sigma), \\
&& \tilde{S}^r(\sigma+2\pi) = e^{\pi iwqR}\tilde{S}^r(\sigma), \nonumber \\
&& \tilde{\lambda}(\sigma+2\pi) = e^{-2\pi iwA^I_YR}\tilde{\lambda}(\sigma). \nonumber 
\end{eqnarray}
By setting $qR=1$ and choosing an appropriate Wilson line, one can see that this theory is 
equivalent to the orbifold mentioned above (winding sectors with odd $w$ form the twisted 
sector). 
Note that the radius of the $S^1$ of the orbifold is twice as large as the original one. 
Similarly, one can also show the equivalence between the KK-Melvin and the orbifold even when 
$qR$ is a rational number. 
Thus, according to the adiabatic argument, the Type I-heterotic duality also holds for this 
case. 
The equivalence to an orbifold for any rational $qR$ is shown in terms of partition 
functions in Type IIA 
case \cite{TU1}. 

When $qR$ is an irrational number, the corresponding worldsheet theory has infinite number 

of twisted sectors, and the equivalence would not make sense. 
However, naively it could be expected that the equivalence persists for any irrational $qR$. 
Suppose that strings live in a spacetime of the form 
$X^{1,6}\times {\bf R}\times {\bf R}^2$, where $X^{1,6}$ is a seven-dimensional spacetime. 
Denote this theory as $C$. 
Then the string theory on $X^{1,6}\times$Melvin can be reinterpreted as the orbifold 
$C/TS$, where $T$ is the translation along the ${\bf R}$ and $S$ is the spatial 
rotation in the ${\bf R}^2$. 
The action of $TS$ does not have any fixed point, and therefore the duality relation might 
exist for general value of $qR$. 

\vspace{2mm}

Once the Type I-heterotic duality is assumed even when the theories live on the KK-Melvin, 
all the duality relations generated by it and the T-duality 
could be used to study the theories, according to the arguments in \cite{unify}. 
We will show below that the use of the duality relations among string theories on the 
KK-Melvin 
is powerful enough to discuss what happens when closed string tachyons condense in the 
heterotic theory on the KK-Melvin.

\vspace{1cm}

\section{Perturbative spectrum} \label{spectrum}

\vspace{5mm}

In this section, we briefly show the perturbative spectrum of the heterotic theory on the 
KK-Melvin with a Wilson line. 
For definiteness, we choose the parameters as $qR=1$ and 
\begin{equation}
A^1_YR = (1,0,\cdots,0).
    \label{Wilsonline}
\end{equation}
This background gives the same spectrum as the non-supersymmetric $SO(32)$ heterotic theory 
in the limit $R\to \infty$ \cite{Suyama1}\cite{Suyama2}. 
Note that in this limit, the adiabatic argument would not be applicable and investigations 
of the tachyon condensation in the theory based on duality arguments might not make sense. 
However, as we will see below, tachyons already appear for small but finite radius of the 
$S^1$ in the KK-Melvin, 
and hence the investigations of the tachyon condensation in the KK-Melvin can be carried out 
by using 
duality arguments. 

To see the spectrum, it is convenient to employ RNS formalism for right-moving fermions. 
Their worldsheet action is 
\begin{equation}
S_R = \frac i{\pi}\int d^2\sigma\ \left\{ \psi_\mu\partial_+\psi^\mu
        +\psi^\dag(\partial_++iq\partial_+Y)\psi+\psi^Y\partial_+\psi^Y \right\}. 
\end{equation}
The choice of the parameters above makes the perturbative calculations very easy. 
It is because the boundary conditions for the winding sectors do not change, while the 
GSO projections for both left- and right-movers are reversed. 
The mass operator and the level-matching condition are 
\begin{eqnarray}
&& \frac{\alpha'}2M^2 = \frac{\alpha'}2p^2+\frac{(Rw)^2}{a\alpha'}+N_L+N_R-E_0, \\
&& N_L-N_R = -wpR, 
\end{eqnarray}
where $N_{L(R)}$ are the ordinary level operators and $E_0$ is the zero-point energy which is 
the same value as the flat background. 
The zeromode $p$ of $Y$ is expressed as follows, 
\begin{eqnarray}
p &=& \frac1R\left[m-\int_0^{2\pi}d\sigma\left\{
  \frac i{4\pi\alpha'}(\partial_\tau\tilde{X}^\dag\tilde{X}
       -\tilde{X}^\dag\partial_\tau\tilde{X})
  -\frac1{2\pi}\psi^\dag\psi \right\} \right. \nonumber \\
  & & \left. -\int_0^{2\pi}d\sigma\frac1{2\pi}{\lambda^1}^\dag\lambda^1 \right]. 
\end{eqnarray}
Here $\tilde{X}$ is defined by the eq.(\ref{tilde}). 

In the secotr with $w=0$, the spectrum is the same as that for the flat background. 
This is because, in this sector, the only difference from the flat case is the modification 
of $p$, and $pR-m$ is an integer for each state in the Fock space. 
Thus the shift of $p$ due to the presence of the nontrivial background can be cancelled by 
shifting the KK-momentum number $m$. 

In the winding sectors with odd $w$, the GSO projections for both the left- and the 
right-movers are reversed. 
One can see that tachyonic states exist only in the NS-NS sector. 
The lightest states in the sector with an odd $w$ are 
\begin{eqnarray}
&& \tilde{\lambda}^k_{-\frac12}|0\rangle_{NS-NS}, \hspace{1cm} 
  \tilde{\lambda}^{k\dag}_{-\frac12}|0\rangle_{NS-NS}, \hspace{5mm} (k=2,\cdots,16), 
     \nonumber \\
&& \tilde{\lambda}^1_{-\frac12}|m=-1\rangle_{NS-NS}, \hspace{1cm}
  \tilde{\lambda}^{1\dag}_{-\frac12}|m=1\rangle_{NS-NS}, 
      \label{tachyons}
\end{eqnarray}
and their masses are
\begin{equation}
\frac{\alpha'}2M^2 = \frac{(wR)^2}{2\alpha'}-1. 
\end{equation}
Therefore, tachyonic states appear in the theory if $R<\sqrt{2\alpha'}$. 
It is clear from the expressions of the states (\ref{tachyons}) that the tachyons couple to 
the gauge fields. 

To see that the above 32 tachyonic states form a vector representation of the $SO(32)$ gauge 
group, it is convenient to consider the bosonic construction of the current algebra. 
For the supersymmetric $SO(32)$ theory, the internal momentum lattice is denoted as 
$\Gamma_{16}$, which is generated by the root lattice of $SO(32)$ and a weight 
\begin{equation}
\left(\frac12,\frac12,\cdots,\frac12\right), 
\end{equation}
in a suitable basis. 
This means that the lattice contains the root lattice and one spinor lattice of $SO(32)$. 
The momentum lattice is shifted by turning on a Wilson line. 
Consider the Wilson line (\ref{Wilsonline}) used in the above calculation. 
Then the shifted lattice contains the vector lattice and one spinor lattice which has the 
opposite chirality to the one in the original lattice $\Gamma_{16}$. 
States in the winding sectors with odd $w$ correspond to weights in the shifted lattice, 
and hence the tachyonic states form the vector representation. 

Note that the tachyons with the winding number $w=+1$ and $w=-1$ have the same mass squared 
and they appear in the same range of $R$, i.e. $R<\sqrt{2\alpha'}$. 
As $R$ becomes smaller, there appear more number of tachyons and the process of the tachyon 
condensation might become more complicated. 
We will restrict ourselves to the simplest case in which tachyons only appear in the 
$w=\pm1$ sectors.

\vspace{1cm}

\section{M-theory dual} \label{M}

\vspace{5mm}

It is well-known \cite{various} that the heterotic theory compactified on $T^3$ is dual to 
M-theory compactified on K3. 
In this section, we assume that the duality between the heterotic theory on 
$T^3\times$Melvin and M-theory on K3$\times$Melvin holds, 
and we discuss the dual picture of the tachyonic instability 
of the heterotic theory. 
Similar dualities are discussed recently \cite{SUGRAdual}. 

As shown in the previous section, the tachyons in the heterotic theory have the winding 
number and the charges which couple to the gauge fields. 
One can find the dual states in M-theory by looking for states which have the corresponding 
quantum numbers. 
From the correspondence of spacetime fields between two supergravity theories, the quantum 
numbers are related as follows \cite{various}. 

The heterotic theory on $T^3$ has the gauge group of the rank 22. 
For generic points of the moduli space, the gauge group is broken to $U(1)^{22}$. 
The corresponding gauge fields in M-theory come from the 3-form field via 
the Kaluza-Klein reduction. 
The number of the gauge fields is the same since the dimension of the second cohomology 
group $H^2(K3)$ is 22. 
Therefore the charged states in the heterotic theory would correspond to bound states which 
include M2-branes wrapped on some 2-cycles in the K3. 
Non-Abelian gauge groups appear when the K3 becomes singular \cite{various}. 
The extra gauge fields (``W-bosons'') come from M2-branes wrapped on the vanishing 2-cycles 
at the singularity. 

The NS-NS B-field couples to the winding number corresponding to the $S^1$ in the KK-Melvin. 
Its dual field in M-theory is the Hodge dual of the 3-form field. 
Therefore the winding states in the heterotic theory would correspond to bound states which 
include M5-branes wrapped on the K3 and the $S^1$ in the KK-Melvin. 

Thus we conclude that the charged tachyons in the heterotic theory on the KK-Melvin would 
correspond to M2-M5 bound states in M-theory. 
Note that the instability we will discuss below is the one due to the bound states of 
branes, and hence the phenomena induced by them might be different from the ones induced by 
tachyonic strings argued in several papers 
\cite{APS}\cite{Vafa}\cite{c-Th}\cite{DV}. 

\vspace{2mm}

It is convenient to see this situation in Type IIA picture, by the dimensional reduction 
along the $S^1$ in the KK-Melvin. 
This is because in this picture the instability can be understood in terms of the local 
physics, 
while in M-theory picture the instability originates in the global geometry. 
The resulting background in the Type IIA picture is known as a fluxbrane, 
\begin{eqnarray}
ds^2 &=& \sqrt{1+q^2r^2}(\eta_{\alpha\beta}dx^\alpha dx^\beta+ds^2_{K3}+dr^2)
        +\frac{r^2}{\sqrt{1+q^2r^2}}d\theta^2, 
           \nonumber \\
e^{\frac43\Phi} &=& 1+q^2r^2, 
      \label{flux}  \\
A_\theta &=& \frac{qr^2}{1+q^2r^2},  \nonumber 
\end{eqnarray}
where $\alpha,\beta=0,\cdots,3$. 
In this picture, the corresponding states to the charged tachyons are D2-D4 bound states 
wrapped on the K3. 

It is disussed \cite{flux3}\cite{flux5} that in the presence of the fluxbrane D4-branes are 
blown up to a 
D6-brane whose worldvolume is $S^2\times$(worldvolume of D4-branes). 
This phenomenon can be understood in terms of the dielectric effect of D-branes \cite{Myers}. 
Suppose, for simplicity, that the D4-branes are at the center of the fluxbrane, i.e. at 
$r=0$. 
Then, around the D4-branes, the background is approximated by the geometry of the flat 
spacetime times K3, the 
constant dilaton, and the constant R-R field strength. 
In particular, the field strength is magnetic, so that it is equivalent to the 8-form 
electric R-R field strength. 
The non-Abelian version of the Chern-Simons terms in the worldvolume theory of D4-branes 
contains the coupling of the D4-branes to the 7-form R-R field. 
This interaction makes the spherical distribution of the D4-branes more stable, and the 
configuration locally has the D6-brane charge (although there is no net D6-brane charge). 
The solution for the dielectric D4-branes in the supergravity is obtained and its 
stability is examined \cite{flux3}\cite{flux5}. 

We would like to relate the tachyonic instability in the heterotic theory to the instability 
of the D2-D4 systems in the fluxbrane against the blowing-up to spherical D2-D4-D6 systems. 
It is based on the following observations. 

First of all, in the heterotic theory the tachyons exist only in the winding sectors, as 
shown in the previous section. 
The duality between the heterotic theory and M-theory would relate the winding number and 
the D4-brane charge. 
Thus it would be natural to relate the tachyonic instability to the one due to the presence 
of the D4-branes. 

In \cite{CostaGutperle}, it is conjectured that Type 0A theory is dual to Type IIA theory 
with the fluxbrane background (\ref{flux}) with the $ds^2_{K3}$ replaced with the flat one. 
In addition, it is argued that the tachyonic instablity in Type 0A theory is related to the 
instability due to pair creations of D6-branes. 
See also \cite{instanton}\cite{instanton2}. 
Through the pair creation, it is argued that the fluxbrane would decay into the flat 
spacetime, so that in our case it would be expected that the blowing-up of the D2-D4 systems 
leads to a stable vacuum. 

\vspace{2mm}

\begin{figure}[htb]
 \epsfxsize=20em
 \centerline{\epsfbox{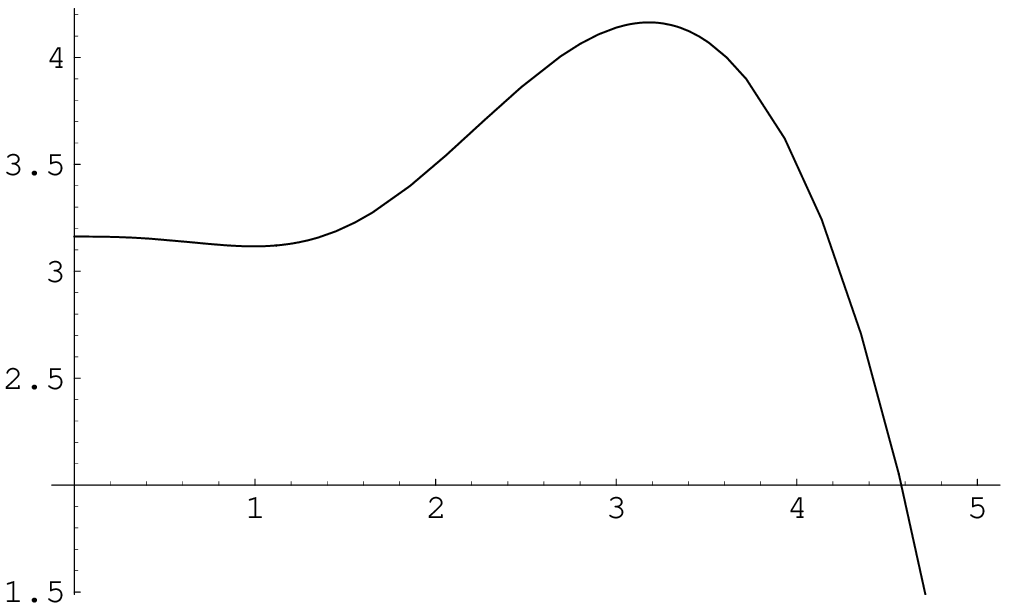}}
 \caption{The potential for small $b$}
   \label{fig1}
 \epsfxsize=20em
 \centerline{\epsfbox{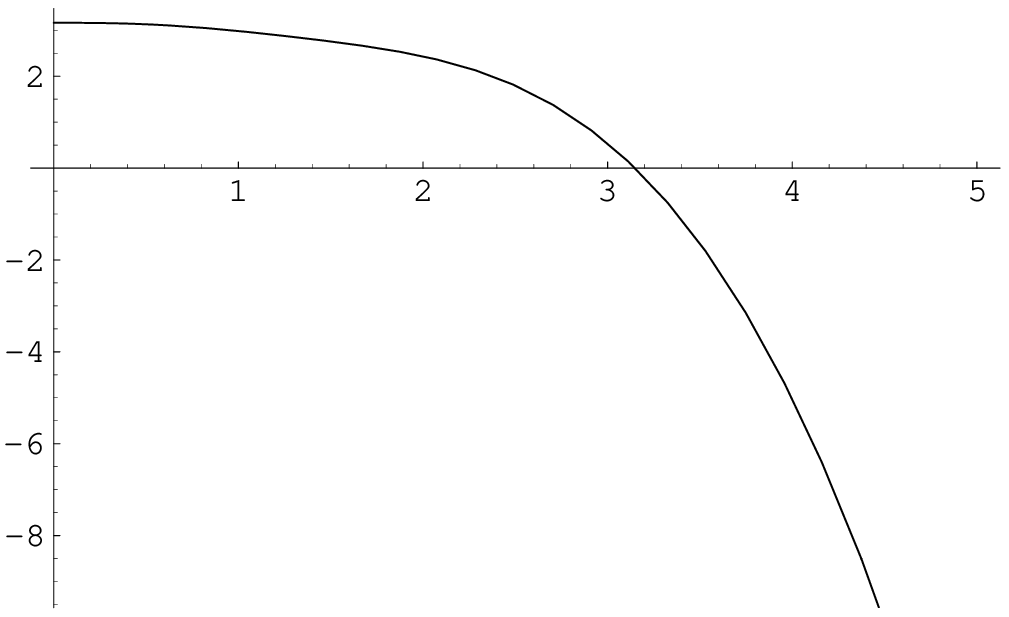}}
 \caption{The potential for large $b$}
   \label{fig2}
\end{figure}

It is very interesting to investigate the potential for the radius $\rho$ of the spherical 
D6-brane in the R-R background. 
The investigations of the gravity solution show that qualitative properties of the spherical 
D6-D4 system 
can be seen from the probe analysis \cite{Myers}\cite{flux5}, although the 
calculations have been 
performed in the flat background geometry which is not consistent. 
The potential is 
\begin{equation}
V(\rho) \propto \sqrt{\rho^4+a}-b\rho^3, 
\end{equation}
where $a,b$ are constants and $b$ is proportional to $q$. 
The shape of the potential $V(\rho)$ is shown in figure \ref{fig1},\ref{fig2}. 

For large $b$, the potential has a unique extremum, i.e. maximum at $\rho=0$ (figure 
\ref{fig1}). 
On the other hand, for small $b$ there is a local minimum at a finite $\rho$ (figure 
\ref{fig2}), and 
there is a potential barrier for the D6-brane to infinitely expand. 
It could be expected that $\rho\to\infty$ corresponds to the decay of the fluxbrane (or, in 
other words, the stabilization of the tachyonic instability). 
Consider the case with fixed $qR$ and vary $R$. 
Then $q$ is inversely proportional to the radius $R$ of the $S^1$ in the KK-Melvin, and one 
finds that the large $b$ case corresponds to the small $R$ and vice versa. 
Therefore there seems to be a nice correspondence between the shape of the potential and the 
pattern of the appearance of the tachyons in the heterotic theory; for large $R$ there is a 
potential barrier to stabilize the vacuum, corresponding to the absence of the tachyon, 
while for small $R$ no obstruction exists toward 
the decay, and thus the instability could be already seen in the perturbative level. 

\vspace{2mm}

Now it is easy to deduce the endpoint of the stabilization of the M-theory which 
would be dual to the heterotic theory on the KK-Melvin. 
The KK-Melvin would decay into the flat background times $S^1$. 
In view of the Type IIA picture, the R-R flux would be cancelled by the dipoles induced by 
the spherical D6-branes. 
Since the energy of the system would decrease by rolling down the potential, the energy of 
the fluxbrane would also be cancelled. 
(It is pointed out that such arguments might be subtle \cite{twistS1}) 

It is natural to expect that the expanding D6-branes will take the D2-D4 systems to 
infinity since they form bound states. 
In addition, it would be possible that some of the D2-D4 systems remain and even condense, 
since the binding energy of them with the D6-branes is finite and they can be separated from 
the expanding D6-branes. 
In the latter case, the charges coupled to the gauge fields would condense 
in the resulting vacuum. 
We will discuss this situations in the next section. 

Recently, it is discussed in \cite{flux14} that the fluxbrane (\ref{flux}) can be 
constructed from the 
D6-branes and the anti-D6-branes which are placed far apart from each other. 
According to this viewpoint, the decay of the fluxbrane is a direct consequence of the 
annihilation of the D6-branes and the spherical D6-branes. 
This result would strongly suggest that the endpoint of the decay of the fluxbrane is the 
supersymmetric vacuum without the R-R flux, as is conjectured in \cite{CostaGutperle}. 
Thus the stable vacuum of the M-theory discussed above would be 
${\bf R}^{1,5}\times$K3$\times S^1$ on which M-theory is dual to the heterotic theory on 
$T^3\times S^1$.

\vspace{1cm}

\section{String theory duals} \label{string}

\vspace{5mm}

In this section, we consider several heterotic theories and their duals to discuss effects 
of the charged tachyon condensation on the gauge symmetries. 
The background of each heterotic theory is of the form $X\times$Melvin, where 
$X$ is a compact manifold. 
We will argue that in some cases the rank of the gauge group is unchanged by the tachyon 
condensation, as opposed to the naive expectation. 

In the following discussions, we will restrict ourselves to the situations in which the 
gauge charges would condense but the winding number is not. 
One can consider such special situations since there are tachyons with the winding 
number both $+1$ and $-1$. 
Therefore, in other words, we will consider the condensation of composite particles which 
consist of tachyons with total winding number zero. 
We will comment on more general situations in section \ref{discuss}. 

\vspace{2mm}

\vspace{5mm}

(i) No compact space

\vspace{3mm}

Consider the simplest case; the $SO(32)$ heterotic theory on the KK-Melvin with the Wilson 
line (\ref{Wilsonline}). 
Recall that this theory has the gauge group $SO(32)$ and the tachyons in the vector 
representation of the $SO(32)$. 
Therefore, if these tachyons acquire nonzero vevs, one may expect that the gauge symmetry is 
broken down 
and the rank of the gauge group is reduced. 

As discussed in section \ref{dual} and in \cite{Suyama2}, the dual description of this 
theory would be Type I theory on the KK-Melvin with the corresponding Wilson line. 
In the Type I picture, the winding strings around the $S^1$ of the KK-Melvin in the 
heterotic picture correspond to the winding D-strings around the same $S^1$. 
The vector representation of the $SO(32)$ can be constructed from the open strings with one 
end on the D-string and the other end on the D9-branes. 
Since there is no Wilson line of the worldvolume gauge field on the D-string, the fermionic 
degrees of freedom of the open strings obey the anti-periodic boundary condition along the 
$S^1$. 
The Wilson line of the spacetime gauge field does not affect the open strings since they 
have the vector index of the $SO(32)$. 
Therefore the strings have the mode expansion with the half-integral moding, and the vector 
states can be obtained. 
Thus the duality suggests that the states corresponding to the tachyons in the heterotic 
theory are the winding D-strings with open strings stretched between them and the D9-branes. 

Now we can consider the condensation of the tachyonic states. 
We have assumed that the winding number is totally cancelled. 
This means that there are the same number of the D-strings and the anti-D-strings. 
Thus they are pair-annihilated and there remain the fundamental open strings. 
However, because of the absence of the D-strings, both ends of the fundamental open strings 
must 
be attached to the D9-branes. 
The resulting strings need not be attached to a single D9-brane, so there are 
still charged states, but they are now in the adjoint representation of the $SO(32)$. 
Therefore we conclude that the endpoint of the charged tachyon condensation with no net 
winding number would be the supersymmetric Type I background with a nonzero vev of 
adjoint fields. 
One can easily see that the only adjoint field that can condense is the massless scalar 
$A_y$ which comes from the $y$-component of the gauge field in ten dimensions. 
The generic vev of the $A_y$ breaks the gauge group to $U(1)^{16}$ and the rank remains the 
same. 
This result is reasonable since the rank of the gauge group must be sixteen to ensure the 
anomaly cancellation in ten dimensions in the Type I theory. 

The above phenomenon can be understood in terms of the low energy field theory. 
From the assumption that the net winding number vanishes, the tachyons in the vector 
representation would form pairs with other tachyons which have the opposite winding number. 
Then the pairs would be in the adjoint representation. 
Hence the fields which have nonzero vevs would be in the adjoint, not in the vector. 
The above Type I picture shows the mechanism explicitly.

\vspace{5mm}

(ii) $X=T^2$

\vspace{3mm}

By compactifying the heterotic theory discussed above on $T^2$, one can find other 
situations with more 
general gauge groups and representations of tachyons, depending on the point of the moduli 
space of the $T^2$ compactification. 
The dual theory would be F-theory \cite{Fth} compactified on K3$\times$Melvin.
\footnote{I would like to thank T.Tani for pointing out this dual picture.}

Recall the duality relation between F-theory on K3 and string theories in eight dimensions. 
If the K3 is the orbifold $T^4/{\bf Z}_2$, F-theory compactified on the K3 can be described 
by a perturbative Type IIB orientifold \cite{F}. 
In this picture, Type IIB theory is compactified on $T^2/{\bf Z}_2$, and one 
orientifold plane and four D7-branes exist on each fixed point. 
This orientifold is just the T-dual of Type I theory compactified on $T^2$ with the Wilson 
lines
\begin{eqnarray}
&& A_5^IR_5 = \left(\mbox{$\frac12$}^4,\mbox{$\frac12$}^4,0^4,0^4\right), \\
&& A_6^IR_6 = \left(\mbox{$\frac12$}^4,0^4,\mbox{$\frac12$}^4,0^4\right),
\end{eqnarray}
where $x^5,x^6$ are the coordinates of the $T^2$. 
This theory is also dual to the heterotic theory on $T^2$. 

Consider the heterotic theory on $T^2\times$Melvin with the above Wilson lines. 
The gauge group of the theory is $SO(8)^4$. 
For each $SO(8)$, there are tachyons in the vector representation. 
The dual theory would be the Type IIB orientifold whose target space is 
$T^2/{\bf Z}_2\times$Melvin. 
The states corresponding to the heterotic tachyons can be deduced as follows. 
The winding strings around the $S^1$ of the KK-Melvin are mapped to D3-branes wrapped on the 
same $S^1$ and the $T^2/{\bf Z}_2$. 
The D3-branes can have the charges for the gauge fields on the D7-branes, which come from 
the open strings stretched between 
them and the D7-branes, as in the previous case. 
One can show that the quantization of the strings produces the vector states. 
Hence we could expect that these D3-brane states are the counterpart of the heterotic 
tachyons. 

The condensation of such states can be discussed in the similar manner to the previous case. 
By assumption, the D3-branes are completely pair-annihilated, and the remaining open strings 
must be attached to the D7-branes at the fixed points. 
Therefore, in the end some adjoint fields living on the D7-branes would acquire nonzero vevs. 
The candidates for the adjoint fields are three massless scalars $A_i$ on the D7-branes. 
The generic vevs of the three adjoint fields can break the gauge symmetry completely. 
However, all the vevs does not correspond to the vacuum configurations. 
Since it is expected that there are sixteen supercharges at the endpoint of the decay, the 
potential for $A_i$ 
is completely 
determined 
\begin{equation}
V = -\frac14\mbox{Tr}[A_i,A_j]^2.
\end{equation}
To obtain the vacuum configuration, all $A_i$'s must commute with each other. 
Then the gauge group is in general broken to its maximal Abelian subgroup, i.e. the rank is 
preserved. 

\vspace{2mm}

In general, the dual of the heterotic theory on $T^2\times$Melvin would be F-theory 
compactified on K3$\times$Melvin. 
The K3 is an elliptic fibration over ${\bf P}^1$. 
This theory would have a description in terms of Type IIB theory, although it is not a 
perturbative one in general. 
The target space is the KK-Melvin times the base space ${\bf P}^1$ of the K3, and the 
modulus of the fiber 
corresponds to the linear combination of the dilaton and the R-R 0-form at each point of the 
base space. 
The fiber degenerates at 24 points at which there is a D7-brane (or in general a (p,q) 
7-brane). 

The winding number in the heterotic theory would be mapped to D3-branes wrapped on the 
${\bf P}^1\times S^1$. 
This is true at the orientifold limit discussed above, since $T^2/{\bf Z}_2$ is 
topologically equivalent to ${\bf P}^1$, and this correspondence would still hold after the 
background is deformed continuously. 
The heterotic tachyons would correspond to the D3-branes with open (p,q)-strings, in 
general, and after the annihilation of the D3-branes the open strings are attached to the 
7-branes. 
Such string configuration produces adjoint representations of 
the gauge groups \cite{junction}.

\vspace{5mm}

(iii) $X=T^4$

\vspace{3mm}

It seems interesting to consider this case, since in the dual picture the condensation of 
D-branes should occur. 
The condensation of D-branes have been discussed in the context of topology-changing 
transitions of the target space, for example, the conifold transitions \cite{conifold}. 

The dual theory would be Type IIA theory on K3$\times$Melvin. 
We are interested in situations in which there are non-Abelian gauge symmetries, and 
hence we consider a singular K3. 
The winding number would be mapped to NS5-branes wrapped on the K3 and the $S^1$ of the 
KK-Melvin. 
In the Type IIA picture, the spacetime gauge field comes from the R-R fields, and hence 
the gauge charges would be provided by D2-branes wrapped on 2-cycles of the K3. 
Thus the corresponding states to the heterotic tachyons would be bound states of the 
NS5-branes and the D2-branes. 

The condensation of the charges corresponds to the condensation of the D2-branes. 
In order for the charged states to have a nonzero vev, there must be massless states coming 
from the D2-branes. 
This can be realized when the D2-branes are wrapped on vanishing cycles at a 
singularity of the K3. 
The massless states are the vector multiplet. 
Thus it is the adjoint states that acquire nonzero vevs, as in the previous cases. 
The supersymmetry at the endpoint of the decay prevent the rank of the gauge group from 
reducing.

\vspace{5mm}

(iv) $X=\mbox{K3}\times T^2$

\vspace{3mm}

We have discussed the situations which are expected to have sixteen supercharges at the end 
of the condensation, and thus what happens after 
the charged tachyon condensation is severely restricted by the supersymmetry. 
Therefore, it would be natural to expect that something different may occur if there are 
less number of supercharges recovered after the condensation. 

In this case, the dual theory would be Type IIA theory on a Calabi-Yau threefold times the 
KK-Melvin. 
The threefold is a K3 fibration over ${\bf P}^1$. 
Non-Abelian gauge symmetries appear when all the fibers become singular simultaneously
\cite{Aspinwall}. 
In general, the condensation of charged states breaks the gauge symmetry and often reduces 
the rank. 
Interestingly enough, it is argued that there are geometric transitions of the threefold 
\cite{Aspinwall} which reduce the rank of the gauge group. 
These transitions are realized as deformations of the complex structure of the singular 
threefold. 
The deformations of the complex structure in general remove, in some sense, vanishing 
cycles, and the homology can change, in contrast to the blowing-up of singularities. 
Therefore, such deformations would be appropriate to describe the condensation of charged 
states. 

The correspondence between the deformations of the complex structure and relevant 
deformations of a theory also seems natural from the worldsheet point of view. 
Consider the string propagation in the vicinity of the singularity in the K3 fiber. 
This is approximated by the string theory on the ALE space, which has a description in terms 
of the Landau-Ginzburg model \cite{ALE}; for $A_{n-1}$ singularity, the superpotential is 
\begin{eqnarray}
&& W(w,x,y,z) = w^{-n}+f(x,y,z), \\
&& f(x,y,z) = x^2+y^2+z^n,
\end{eqnarray}
where $f(x,y,z)=0$ defines the $A_{n-1}$ singularity. 
The deformations of the complex structure correspond to adding polynomials to 
$f(x,y,z)$, and hence they change the superpotential. 
On the other hand, the fixed point of the RG flow is determined by the superpotential 
\cite{wsRG}\cite{wsRG2}. 
Therefore, the deformations of $f(x,y,z)$ would change the RG behavior, which is an 
appropriate property for the relevant deformations of the theory. 

It might be possible that the tachyon condensation for $X=T^4$ case could be understood in 
terms of the background geometry in the dual Type IIA picture. 
For the K3 manifold, the deformations of the complex structure are equivalent to the 
deformations of the K\"ahler structure. 
Thus all deformations of singularities can be reinterpreted as the blowing-up, and hence 
the rank of the gauge symmetry in the Type IIA picture is not reduced.

\vspace{1cm}

\section{Discussion} \label{discuss}

\vspace{5mm}

We have discussed the charged tachyon condensation in the heterotic theory on the KK-Melvin, 
and focused on whether (and how) the gauge symmetry breaking occurs. 
The discussions are based on the assumption that the adiabatic argument can be applied to 
our non-supersymmetric setting. 
Once we assume that it is the case, we can argue the charged tachyon condensation in terms 
of the various dual pictures. 
The gravitational dual strongly suggests that the supersymmetry-breaking background would 
decay into a supersymmetric vacuum, and the gauge charges could condense. 
To see the effect of the condensation of the charges on the gauge symmetry, we have 
investigated string theory duals and argue the fate of the gauge group, in particular of its 
rank. 

We have excluded the situations in which there is a net winding number after the 
condensation. 
One could argue that the investigations of such situations are not necessary as follows: 
After the condensation, any winding states would turn into massive states. 
It can be deduced from the fact that in every dual picture the winding number is mapped to 
a number of an extended object with finite volume. 
It might be possible for the winding states to be massless if the corresponding $S^1$ 
shrinks down (or decompactifies in the T-dual picture) after the condensation. 
However, it is suggested that the condensation deforms the background in the opposite 
direction \cite{Suyama3}\cite{flux14}. 
Since massive states would not have nonzero vevs, the condensation with nonvanishing winding 
number would not occur. 

We have not discussed any quantitative properties of the condensation. 
It is mainly because there is no preserved supersymmetry at all, during the condensation. 
Recently, it is pointed out that the worldsheet supersymmetry is a powerful tool to 
consider closed string tachyons in non-supersymmetric backgrounds 
\cite{Vafa}\cite{c-Th}\cite{twistS1}, 
and moreover, the spacetime action of the tachyons are proposed \cite{DV}. 
It is very interesting to apply such techniques to the heterotic cases. 
Note that in such research the perturbations which preserve the worldsheet supersymmetry 
is considered. 
However, it would be important to discuss more general perturbations, although it is much 
harder to handle. 

An interesting viewpoint obtained in our analysis is that the tachyon condensation in the 
heterotic theory might correspond to a geometric deformation of the background in the dual 
Type II theory. 
This would imply that the IR dynamics of such a theory might be quite different from its UV 
dynamics. 
It will be interesting to construct such examples explicitly. 

In the case of the open string tachyon condensation, the phenomena can be interpreted as 
decays of some unstable D-brane systems. 
It seems that in the case of closed string tachyon condensation there is no such a unified 
understanding of the phenomena. 
Fortunately, many non-supersymmetric backgrounds are known to have tachyons, and hence 
it will be important to understand them in a unified way.

\vspace{1cm}

{\Large {\bf Acknowledgements}}

\vspace{5mm}

I would like to thank M. Natsuume, Y. Sato and T. Tani for valuable discussions.


\vspace{1.5cm}





















\newpage


\begin{thebibliography}{99}

\bibitem{open}
A.Sen, 
{\it Stable Non-BPS States in String Theory}, 
JHEP {\bf 9806} (1998) 007, hep-th/9803194. 

\bibitem{open2}
K.Ohmori, 
{\it A Review on Tachyon Condensation in Open String Field Theories}, 
hep-th/0102085. 

\bibitem{CostaGutperle}
M.S.Costa, M.Gutperle, 
{\it The Kaluza-Klein Melvin Solution in M-theory}, 
JHEP {\bf 0103} (2001) 027, hep-th/0012072. 

\bibitem{RussoTseytlin}
J.G.Russo, A.A.Tseytlin, 
{\it Exactly solvable string models of curved space-time}, 
Nucl. Phys. {\bf B449} (1995) 91, hep-th/9502038;  
{\it Magnetic flux tube models in superstring theory}, 
Nucl. Phys. {\bf B461} (1996) 131, hep-th/9508068.  

\bibitem{instanton}
F.Dowker, J.P.Gauntlett, D.A.Kastor, J.Traschen, 
{\it Pair Creation of Dilaton Black Holes}, 
Phys. Rev. {\bf D49} (1994) 2909, hep-th/9309075. 

\bibitem{instanton2}
F.Dowker, J.P.Gauntlett, S.B.Giddings, G.T.Horowitz, 
{\it On Pair Creation of Extremal Black Holes and Kaluza-Klein Monopoles}, 
Phys. Rev. {\bf D50} (1994) 2662, hep-th/9312172; \ 
{\it The Decay of Magnetic Fields in Kaluza-Klein Theory}, 
Phys. Rev {\bf D52} (1995) 6929, hep-th/9507143; \ 
{\it Nucleation of P-Branes and Fundamental Strings}, 
Phys. Rev. {\bf D53} (1996) 7115, hep-th/9512154. 

\bibitem{Suyama1}
T.Suyama, 
{\it Closed String Tachyons in Non-supersymmetric Heterotic Theories}, 
JHEP {\bf 0108} (2001) 037, hep-th/0106079. 

\bibitem{Suyama2}
T.Suyama, 
{\it Melvin Background in Heterotic Theories}, 
hep-th/0107116, to appear in Nucl. Phys. B.
 
\bibitem{flux-1}
D.V.Gal'tsov, O.A.Rytchkov, 
{\it Generating branes via sigma-models}, 
Phys. Rev. {\bf D58} (1998) 122001, hep-th/9801160.

\bibitem{flux0}
C.-M.Chen, D.V.Gal'tsov and S.A.Sharakin, 
{\it Intersecting $M$-fluxbranes},
Grav. Cosmol. {\bf 5} (1999) 45, hep-th/9908132.

\bibitem{flux1}
P. M.Saffin, 
{\it Gravitating Fluxbranes}, 
Phys. Rev. {\bf D64} (2001) 024014, 
gr-qc/0104014. 

\bibitem{flux2}
M.Gutperle, A.Strominger, 
{\it Fluxbranes in String Theory}, 
JHEP {\bf 0106} (2001) 035, hep-th/0104136. 

\bibitem{flux3}
M.S.Costa, C.A.R.Herdeiro, L.Cornalba, 
{\it Flux-branes and the Dielectric Effect in String Theory}, 
hep-th/0105023.

\bibitem{flux4}
R.Emparan, 
{\it Tubular Branes in Fluxbranes}, 
Nucl. Phys. {\bf B610} (2001) 169, hep-th/0105062. 

\bibitem{flux5}
D.Brecher, P.M.Saffin, 
{\it A note on the Supergravity Description of Dielectric Branes}, 
Nucl. Phys. {\bf B613} (2001) 218, hep-th/0106206. 

\bibitem{flux6}
L.Motl, 
{\it Melvin Matrix Models}, 
hep-th/0107002. 

\bibitem{flux7}
A.M.Uranga, 
{\it Wrapped fluxbranes}, 
hep-th/0108196. 

\bibitem{Suyama3}
T.Suyama, 
{\it Properties of String Theory on Kaluza-Klein Melvin Background}, 
hep-th/0110077. 

\bibitem{flux9}
J.G.Russo, A.A.Tseytlin, 
{\it Supersymmetric fluxbrane intersections and closed string tachyons}, 
hep-th/0110107. 

\bibitem{flux10}
C.-M.Chen, D.V.Gal'tsov, P.M.Saffin, 
{\it Supergravity Fluxbranes in Various Dimensions}, 
hep-th/0110164. 

\bibitem{flux11}
J.Figueroa-O'Farrill, J.Simon, 
{\it Generalised supersymmetric fluxbranes}, 
hep-th/0110170. 

\bibitem{flux12}
E.Dudas, J.Mourad, 
{\it D-branes in String theory Melvin backgrounds}, 
hep-th/0110186. 

\bibitem{TU1}
T.Takayanagi, T.Uesugi, 
{\it Orbifolds as Melvin Geometry}, 
hep-th/0110099. 

\bibitem{TU2}
T.Takayanagi, T.Uesugi, 
{\it D-branes in Melvin Background}, 
hep-th/0110200. 

\bibitem{flux14}
R.Emparan, M.Gutperle, 
{\it From p-branes to fluxbranes and back}, 
hep-th/0111177. 

\bibitem{flux15}
Y.Michishita, P.Yi, 
{\it D-Brane Probe and Closed String Tachyons}, 
hep-th/0111199. 

\bibitem{twistS1}
J.R.David, M.Gutperle, M.Headrick, S.Minwalla, 
{\it Closed String Tachyon Condensation on Twisted Circles}, 
hep-th/0111212.

\bibitem{APS}
A.Adams, J.Polchinski, E.Silverstein, 
{\it Don't Panic! Closed String Tachyons in ALE Spacetimes}, 
JHEP {\bf 0110} (2001) 029, hep-th/0108075. 

\bibitem{Vafa}

C.Vafa, 
{\it Mirror Symmetry and Closed String Tachyon Condensation}, 
hep-th/0111051. 

\bibitem{c-Th}
J.A.Harvey, D.Kutasov, E.J.Martinec, G.Moore, 
{\it Localized Tachyons and RG Flows}, 
hep-th/0111154. 

\bibitem{DV}
A.Dabholkar, C.Vafa, 
{\it tt* Geometry and Closed String Tachyon Potential}, 
hep-th/0111155. 

\bibitem{adiabatic}
C.Vafa, E.Witten, 
{\it Dual String Pairs With N=1 And N=2 Supersymmetry In Four Dimensions}, 
Nucl. Phys. Proc. Suppl. {\bf 46} (1996) 225, hep-th/9507050. 

\bibitem{Myers}
R.C.Myers, 
{\it Dielectric-Branes}, 
JHEP {\bf 9912} (1999) 022, hep-th/9910053. 

\bibitem{Pol}
J. Polchinski, 
{\it String Theory I \& II}, 
Cambridge University Press. 

\bibitem{lessSUSY}
A.Sen, C.Vafa, 
{\it Dual Pairs of Type II String Compactification}, 
Nucl. Phys. {\bf B455} (1995) 165, hep-th/9508064. 

\bibitem{unify}
A.Sen, 
{\it Unification of String Dualities}, 
Nucl. Phys. Proc. Suppl. {\bf 58} (1997) 5, hep-th/9609176. 

\bibitem{Harvey}
J.A.Harvey, 
{\it String Duality and Non-supersymmetric Strings}, 
Phys. Rev. {\bf D59} (1999) 026002, hep-th/9807213. 

\bibitem{KKS}
S.Kachru, J.Kumar, E.Silverstein, 
{\it Vacuum Energy Cancellation in a Non-supersymmetric String}, 
Phys. Rev. {\bf D59} (1999) 106004, hep-th/9807076. 

\bibitem{various}
E.Witten, 
{\it String Theory Dynamics In Various Dimensions}, 
Nucl. Phys. {\bf B443} (1995) 85, hep-th/9503124. 

\bibitem{SUGRAdual}
K.Behrndt, E.Bergshoeff, D.Roest, P.Sundell, 
{\it Massive Dualities in Six Dimensions}, 
hep-th/0112071. 

\bibitem{Fth}
C.Vafa, 
{\it Evidence for F-Theory}, 
Nucl. Phys. {\bf B469} (1996) 403, hep-th/9602022. 

\bibitem{F}
A.Sen, 
{\it F-theory and Orientifolds}, 
Nucl. Phys. {\bf B475} (1996) 562, hep-th/9605150. 

\bibitem{junction}
O.DeWolfe, B.Zwiebach, 
{\it String Junctions for Arbitrary Lie Algebra Representations}, 
Nucl. Phys. {\bf B541} (1999) 509, hep-th/9804210. 

\bibitem{conifold}
B.R.Greene, D.R.Morrison, A.Strominger, 
{\it Black Hole Condensation and the Unification of String Vacua}, 
Nucl. Phys. {\bf B451} (1995) 109, hep-th/9504145. 

\bibitem{Aspinwall}
P.S.Aspinwall, 
{\it Compactification, Geometry and Duality: N=2}, 
hep-th/0001001. 

\bibitem{ALE}
H.Ooguri, C.Vafa, 
{\it Two-Dimensional Black Hole and Singularities of CY Manifolds}, 
Nucl. Phys. {\bf B463} (1996) 55, hep-th/9511164. 

\bibitem{wsRG}
E.Martinec, 
{\it Algebraic Geometry and Effective Lagrangians}, 
Phys. Lett. {\bf B217} (1989) 431. 

\bibitem{wsRG2}
C.Vafa, N.Warner, 
{\it Catastrophes and the Classification of Conformal Theories}, 
Phys. Lett. {\bf B218} (1989) 51.








\end{thebibliography}
\end{document}